\documentclass[aps,prd,showpacs,amsmath,10pt,superscriptaddress,floatfix,nofootinbib,twocolumn]{revtex4-2}

\usepackage{epsfig,amssymb,amsfonts,amsmath,mathtools,bm,color,xcolor,graphicx,braket,adjustbox,esint,upgreek,dcolumn,psfrag,hyperref}

\usepackage[utf8]{inputenc}
\usepackage[english]{babel}

\newcommand{\be}{\begin{equation}}
\newcommand{\ee}{\end{equation}}
\newcommand{\bea}{\begin{eqnarray}}
\newcommand{\eea}{\end{eqnarray}}
\newcommand{\beas}{\begin{eqnarray*}}
\newcommand{\eeas}{\end{eqnarray*}}
\newcommand{\ds}{\displaystyle}
\newcommand{\vep}{{\bm p}}

\def\vec#1{\boldsymbol{#1}}

\graphicspath{{Figs.dir/}}

\newcommand{\fzj}{\affiliation{Institute for Advanced Simulation, Institut f\"ur Kernphysik and\\ J\"ulich Center for Hadron Physics, Forschungszentrum J\"ulich, D-52425 J\"ulich, Germany}}
 
\newcommand{\ijs}{\affiliation{Jozef Stefan Institute, Jamova 39, 1000, Ljubljana, Slovenia}}

\newcommand{\lis}{\affiliation{CeFEMA, Center of Physics and Engineering of Advanced Materials, Instituto Superior T{\'e}cnico, Av. Rovisco Pais, 1 1049-001 Lisboa, Portugal}}

\begin{document}

\title{Do near-threshold molecular states mix with neighbouring
$\bar QQ$ states?}

\author{Christoph~Hanhart}
\fzj

\author{Alexey Nefediev}
\ijs
\lis

\begin{abstract}
The last two decades are marked by a renaissance in hadronic spectroscopy caused by the arrival of vast experimental information on exotic states in the spectrum of charmonium and bottomonium. Most of such states have properties at odds with the predictions of the quark model and reside very close to strong hadronic thresholds. Prominent examples are provided by the glorious $X(3872)$ charmonium-like state and the doubly charmed tetraquark $T_{cc}^+$ with the masses within less than 1 MeV from the $D\bar{D}^*$ and $DD^*$ open-charm thresholds, respectively. The universality of this feature hints towards the existence of a general pattern for such exotic states. In this work we discuss a possible generic mechanism for the formation of near-threshold molecular states as a result of the strong coupling of compact quark states with a hadronic
continuum channel.
The compact states that survive the strong coupling limit decouple from
the continuum channel and therefore also from the formed hadronic
molecule --- if realised this scenario would provide a justification
to treat hadronic molecules isolated, ignoring the possible influence
from surrounding, compact quark-model states. 
We confront the phenomenology of the $D_{s1}(2460)$ and $D_{s1}(2536)$ 
with this picture and find consistency, although other explanations remain
possible for those states.
\end{abstract}


\maketitle

\section{Introduction}

Recent discoveries in the spectroscopy of hadrons containing heavy quarks resulted in the appearance of a new and fast developing branch of hadronic physics which deals with exotic states. The very notion of exotics emphasises a special status of such states as opposed to the ordinary ones, predicted and well described by the quark model. A considerable progress in this field has been achieved due to the intensive and fruitful experimental studies of the spectrum of heavy quarkonia above the open-flavour threshold, that is, above the energy when decays of the heavy quarkonium to a pair of heavy-light mesons becomes available. The first state in the family --- the charmonium-like state $\chi_{c1}(3872)$ also known as $X(3872)$ --- was discovered by the Belle Collaboration in 2003 \cite{Choi:2003ue}. Since then it attracts a lot of attention of both theorists, in attempts to understand its nature, and experimentalists, who keep on searching for this state in additional decay and production channels and provide more precise data for the already known modes. In general, the progress in the field of exotic hadrons containing heavy quarks achieved in the two decades after the discovery of the $X(3872)$ is tremendous, and already several dozen well established states in the spectrum of charmonium and bottomonium are conventionally qualified as exotic --- see, for example, the recent dedicated reviews~\cite{Esposito:2016noz,Chen:2016qju,Ali:2017jda,Lebed:2016hpi,Olsen:2017bmm,Guo:2017jvc,Brambilla:2019esw}. 

Many exotic charmonium- and bottomonium-like states reside near strong hadronic thresholds. The most prominent examples are provided by the aforementioned $X(3872)$ and the charged state $T_{cc}^+$ discovered recently by the LHCb Collaboration \cite{LHCb:2021vvq}. In both cases the proximity of the resonance to a nearby hadronic threshold ($D^0\bar{D}^{*0}$ and $D^{*+}D^0$, respectively) is remarkable and hints towards the existence of a deep general reason for the pole of the amplitude to approach the strong threshold located near by.
On the other hand there is a striking difference between the $X(3872)$ and the $T_{cc}^+$: The
former shares its quantum numbers with $\bar cc$ states while the latter does not. In other words, while the $X(3872)$ might contain a prominent quark-antiquark component, the $T_{cc}^+$ must contain (at least) four quarks. Thus, a physical picture that claims a
common understanding of both states must also explain why it is not distorted significantly
by the $\bar cc$ states present in the one case but absent in the other. Indeed, in some works the impact of quarkonium states on exotics is studied, and a strong cross talk of the two is observed --- see, for example,~\cite{Kalashnikova:2009gt,Takizawa:2012hy,Ortega:2012rs,Coito:2012vf,Cincioglu:2016fkm} as well as a recent review \cite{Kalashnikova:2018vkv} and the references therein.

In this paper we propose a mechanism based on a strong interplay of
compact poles with a continuum channel, which could explain the appearance of near-threshold molecular states as a general feature of the system rather than a strongly fine-tuned effect for systems with and without $\bar QQ$ states
present. The picture that emerges implies
that, when molecular type states appear in a given two-hadron channel, the $\bar QQ$
states with the same quantum numbers decouple from this two-hadron channel.
Thus it provides an argument that the mixing of molecular states with 
compact quark model states is suppressed.
We also argue that this picture is in line with the large-$N_c$ limit, where infinite
towers of stable $\bar QQ$ states are predicted to exist. It is also consistent with 
the proposed existence of
analogous towers made of four quarks~\cite{Weinberg:2013cfa,Maiani:2018tfe,Lucha:2021mwx} as well as their absence --- this will be 
discussed towards the end of this paper.

 Ideologically this work can be regarded as a further development of the ideas of the interplay between the quark and hadronic degrees of freedom in near-threshold resonances investigated previously: (i) in Ref.~\cite{Baru:2010ww} for one quark compact state and one continuum channel, (ii) in Ref.~\cite{Hanhart:2011jz} for multiple hadronic channels, and in Ref.~\cite{Hammer:2016prh}, where a multi-resonance situation was considered and a collective-like behaviour of some compact states as a result of strong coupled-channel effects was observed. This study develops further
 Ref.~\cite{Hammer:2016prh} as it provides various analytic insights (see, in particular, Secs.~\ref{sec:general} and \ref{residues}) as well as the confrontation of the emerging phenomenology to the case of the $D_{s1}(2460)$ and $D_{s1}(2536)$ mesons.

\section{Weinberg approach and pole counting}
\label{Weinberg}

We start from a brief introduction into the Weinberg approach to establishing the nature of hadronic states from the data \cite{Weinberg:1962hj}, largely following
Ref.~\cite{Guo:2017jvc}. Consider the simplest coupled-channel problem for an elementary (compact, that is, formed by the QCD confining forces --- no particular assumptions on its quark content are required) state $|\psi_0\rangle$ and a $S$-wave two-meson channel $|M_1M_2 \rangle$ described by the two-component wave function,
\be
|\Psi\rangle=\genfrac{(}{)}{0pt}{}{\lambda|\psi_0\rangle}{\chi(p)|M_1M_2 \rangle},
\label{state}
\ee
which obeys a Schr{\"o}dinger equation (the energy is counted from the two-body threshold, $E=M-M_{\rm th}$),
\be
{\cal H}|\Psi\rangle=E|\Psi\rangle,\quad
{\cal H}=
\begin{pmatrix}
E_0&\hat{v}\\
\hat{v}&p^2/(2\mu)
\end{pmatrix},
\label{Scheq}
\ee
where $E_0$ is the bare energy of the compact state, $\vep$ is the relative momentum in the two-body system, $\mu$ is the reduced mass of the mesons $M_1$ and $M_2$, and the off diagonal potential $\hat{v}$ provides transitions between the two components of the wave function (w.f.) defined in Eq.~(\ref{state}),
\be
\langle \psi_0|\hat{v}|M_1M_2 \rangle=\langle M_1M_2|\hat{v}|\psi_0 \rangle=f(\vep).
\label{Vdef}
\ee

In the expression for the Hamiltonian it is used that, via a proper
field redefinition, nonperturbative two-particle interactions
can be absorbed into a pole leaving only perturbative two-particle
interactions\footnote{This holds only if
just a single state resides in the kinematic regime of interest~\cite{Baru:2010ww}.} --- thus in leading order the meson-meson Hamiltonian is
given by the two-meson kinetic energy only.

Naturally, the factor $\lambda$ quantifies the admixture of the two components in
the physical state since
\be
\left|\langle\psi_0|\Psi\rangle\right|^2=\lambda^2
\ee
defines the probability to find the compact component in the physical
wave function.

If there is a bound state in the system with the eigenenergy $E_B$, the Schr\"odinger equation provided in Eq.~(\ref{Scheq}) gives for the two-meson wave function at the pole, $E=-E_B$,
\be
\chi(p) = \lambda\frac{f(\vec{p})}{E_B+p^2/(2\mu)},
\ee
and the normalisation condition for the w.f. (\ref{state}) reads
\bea
\langle \Psi|\Psi\rangle &=& \lambda^2+\int\frac{d^3p}{(2\pi)^3}|\chi(\vep)|^2\nonumber\\[-2mm]
\label{norm}\\[-2mm]
&=&\lambda^2\left\{1+\int\frac{d^3p}{(2\pi)^3}\left(\frac{f(\vec{p})}{E_B+p^2/(2\mu)}\right)^2\right\}=1.\nonumber
\eea
This expression provides a connection between the transition form
factor $f(\vec{p})$ and $\lambda$,
\bea
\frac{1}{\lambda^2}-1&=&
\int\frac{d^3p}{(2\pi)^3}\left(\frac{f(\vep)}{p^2/(2\mu)+E_B}\right)^2\nonumber\\[-2mm]
\label{tgtheta}\\[-2mm]
&=&\frac{\mu^2f^2(0)}{2\pi\gamma}+{\mathcal O}\left(\frac{\gamma}{\beta}\right),\nonumber
\eea
where we introduced the binding momentum $\gamma=\sqrt{2\mu E_B}$, and $\beta$ is the mass 
scale of change of $f(\vec{p})$, which
implies that $f(\vec{p})=f(0)+{\mathcal O}(p/\beta)$.
The size of $\beta$ is determined either by the mass of the lightest exchange particle (in particular, the mass of the pion, if pion exchange is allowed
for the system at hand) or the next threshold (for a detailed discussion of the latter effect see Ref.~\cite{Matuschek:2020gqe}).
For a discussion of possible range corrections to the Weinberg
expressions we refer to the works \cite{Hanhart:2007wa,Song:2022yvz,Albaladejo:2022sux,Kinugawa:2022fzn}.
As one can read off from Eq.~(\ref{norm}), $\lambda^2$ is identical to the
wave function renormalisation constant $Z$. In particular, the zero-binding limit of $E_B\to 0$ implies that $Z\to 0$ --- see also Ref.~\cite{Hyodo:2014bda} where the latter observation is formulated and proved in the form of a theorem.

We therefore get from Eq.~(\ref{tgtheta})
\be
f^2(0) = \frac{2\pi}{\mu}\left(\frac{\gamma}{\mu}\right)\left(\frac{1}{\lambda^2}-1\right).
\label{g2def}
\ee

It is also of interest below to extract within the formalism just outlined the residue of the $T$-matrix at the bound state pole which comes as the unrenormalised coupling $f^2(0)$ multiplied by the renormalisation constant (as explained above, it is identical to $\lambda^2$) and the relativistic normalisation factor,
\bea
g_{\rm eff}^2&=&8m_1m_2(m_1+m_2)\lambda^2f^2(0)\nonumber\\[-2mm]
\label{geffdef}\\[-2mm]
&=&16\pi(m_1+m_2)^2\left(\frac{\gamma}{\mu}\right)\left(1-\lambda^2\right),\nonumber
\eea
where the chosen name for the residue reflects that it
is the parameter that controls the coupling of the molecule to its constituents. 

To connect the quantities discussed above to observables we may take
a step back and evaluate the scattering amplitude in the hadronic
channel, again neglecting range corrections. Then
one finds~\cite{Guo:2017jvc}
\bea
T(k)&=&{f^2(0)}\left(E+E_B+\frac{\mu}{2\pi}f^2(0)(ik+\gamma)\right)^{-1}\nonumber\\[-2mm]
\label{Flatte}\\[-2mm]
&=&-\frac{2\pi}{\mu}\left(a^{-1}+\frac12r_ek^2-ik\right)^{-1},\nonumber
\eea
where the momentum $k$ is defined as
\be
k(E)=\sqrt{2\mu E}\Theta(E)+i\sqrt{-2\mu E}\Theta(-E),
\label{kE}
\ee
with $\Theta$ for the step-like function.
Equation~(\ref{Flatte}) can be easily recognised as the celebrated Flatt{\'e} distribution for a single resonance coupled to an $S$-wave hadronic channel, with the coupling constant
\be
g_f=\frac{\mu}{\pi}f^2(0),
\ee
while the last formula in Eq.~(\ref{Flatte}) is no more than the effective range expansion with the low-energy parameters
\be
a^{-1}=-2E_B/g_f-\gamma,\quad r_e=-2/(\mu g_f),
\label{Fa}
\ee
for the inverse scattering length and the effective range, respectively. 
Employing Eq.~(\ref{g2def}) one finds
\bea
a&=&\frac{2(1-\lambda^2)}{(2-\lambda^2)}\frac{1}{\gamma}+\mathcal{O}\left(\frac{\gamma}{\beta}\right),\nonumber\\[-2mm]
\label{are}\\[-2mm]
r_e&=&-\frac{\lambda^2}{(1-\lambda^2)}\frac{1}{\gamma}+\mathcal{O}\left(\frac{\gamma}{\beta}\right),\nonumber
\eea
so that the information on the nature of a near-threshold resonance can be extracted directly from data. Relations (\ref{are}) imply that the case of a compact state, $\lambda^2\approx 1$, corresponds to a large, negative effective range and small scattering length. Then the amplitude (\ref{Flatte}) possesses two nearly symmetric near-threshold poles. In the opposite limit of the molecular state with $\lambda^2\to 0$, one has a large scattering length and small effective range, so that only one near-threshold pole survives. These conclusions are in line with the pole counting rules proposed in Ref.~\cite{Morgan:1992ge}. From Eq.~(\ref{Fa}) one can easily see that the weak coupling regime ($g_f\to 0$) corresponds to a compact state while the strong coupling regime ($g_f\to \infty$) is compatible with a molecule, in agreement with natural expectations as long as we look at a single, isolated state.
 
In conclusion of these general considerations and before we discuss a particular model to study the properties of a multiresonance system, we would like to comment on some basic notions often referred to in this work. 
First, we call a state ``near-threshold" as soon as the corrections to the Weinberg 
formulae introduced above are small. This implies that $\gamma/\beta\ll 1$ and that the energy dependence of the contribution of the corresponding hadronic channel to the self-energy of the studied resonance, which is proportional to the momentum $k(E)$ [see Eq.~(\ref{kE})], is essential for understanding its properties. 

Second, the proximity of a resonance to a hadronic threshold hints towards but does not automatically guarantee its molecular nature. Indeed, in addition, the resonance must couple to the corresponding hadronic channel sufficiently strongly. 
For example, as follows from Eq.~(\ref{Fa}) above, the coupling $g_f$ needs to be large enough to suppress the effective range $r_e$ and thus move one of the poles of the amplitude (\ref{Flatte}) sufficiently far away from the threshold. This implies that the probability to observe the resonance as a compact object decreases, that is $\lambda^2\to 0$, as it follows from the second formula in Eq.~(\ref{are}). 

All above arguments can be put together to claim that a near-threshold molecular state can be formed in the spectrum only if (i) it resides sufficiently close to the corresponding $S$-wave hadronic threshold and (ii) couples to it sufficiently strongly, so that the amplitude develops a single near-threshold pole which leaves a significant
imprint on observables. As argued in Ref.~\cite{Guo:2014iya} the threshold nonanalyticity alone is 
in general not strong enough for this effect and needs a nearby pole as an amplifier.

\section{General considerations}
\label{sec:general}

In this section we propose a simple coupled-channel model, which allows us to study the trajectories of the
poles in the complex momentum plane and understand the fate of the compact resonance states after they couple to a
single continuum channel. In particular, following Ref.~\cite{Hammer:2016prh} we consider a set of $N$ scalar resonances, $R_n=(\bar{Q}Q)_n$, coupled to a continuum channel $\varphi\bar{\varphi}$ with $\varphi=\bar{q}Q$ ($\bar{\varphi}=\bar{Q}q$) being a scalar (anti)meson, 
\bea 
{\cal L}&=&\frac12\left[(\partial_\mu\varphi)^2-m^2\varphi^2\right]+\frac12\left[(\partial_\mu\bar{\varphi})^2-m^2\bar{\varphi}^2\right]\nonumber\\[-2mm]
\label{lag}\\[-2mm]
&+&\sum_{n=1}^N\left[\frac12(\partial_\mu R_n )^2-\frac12 M_n^2R_n^2+G_nR_n\bar{\varphi}\varphi\right].\nonumber
\eea

The quarks $Q$ and $q$ have the masses $M_Q$ and $m_q$, respectively, and it is assumed that $M_Q\gg m_q$, so that the model (\ref{lag}) mimics interactions of quarkonia with open-flavour channels in the hadronic spectrum of heavy quarks.

The $\varphi\bar{\varphi}$ scattering potential through the resonances $R_n$ takes the form
\be
V(s)=-\sum_{n=1}^N\frac{G_n^2}{s-M_n^2}
=-(2m)^2\sum_{n=1}^N \frac{g_n^2}{s-M_n^2},
\label{pot}
\ee
where we defined the dimensionless couplings $g_n=G_n/(2m)$ to be used in what follows. 

The $\varphi\bar{\varphi}$ scattering amplitude $T(s)$ is a solution of the Lippmann-Schwinger equation,
\be
T(s)=V(s)+V(s)G(s)T(s),
\label{LSe}
\ee
where $G(s)$ denotes the so-called scalar loop of a $\varphi$ and
$\bar \varphi$ propagating from a point-like source to a point-like
sink, often named polarisation operator. Combining Eqs.~(\ref{pot}) and (\ref{LSe})
one finds that 
\bea
T^{-1}(s)&=&V^{-1}(s)-G(s)\nonumber\\[-3mm]
\label{eq1}\\[-3mm]
&=&-\left[4m^2\sum_{n=1}^N \frac{g_n^2}{s-M_n^2}\right]^{-1}-G(s).
\nonumber
\eea
Physical states are defined as the poles of $T(s)$ or, alternatively, the zeros of $T^{-1}(s)$.

It needs to be mentioned that the amplitude given in Eq.~(\ref{eq1}) possesses relatively simple analytical properties, which is a result of a very simple approach used to obtain it. 
Indeed, the potential provided in Eq.~(\ref{pot}) relies entirely on $s$-channel exchanges between stable particles 
$\varphi$ and $\bar{\varphi}$. Thus, in the absence of the $t$- and $u$-channel exchanges and additional branch points related to the decay modes of the particles $\varphi$, one does not encounter subtleties with a sophisticated structure of the amplitude in the energy complex plane such as left-hand cuts, anomalous thresholds, multibody thresholds, and so on. While a more realistic model clearly calls for their consideration, we do not expect that
their presence will change the qualitative behaviour of the amplitude reported in this work.

For further convenience and unless stated otherwise we pass over to the unitary-cut-free complex plane of the three-momentum $k$ by setting
\be
s=4(k^2+m^2).
\ee

Then the equation for the poles reads
\be
1+m^2\left[\sum_{n=1}^N \frac{g_n^2}{k^2-\Delta_n^2}\right]G(k)=0,
\label{eqgen}
\ee
where each quantity $\Delta_n=\frac12\sqrt{M_n^2-4m^2}$ takes either a real or imaginary value, depending on the position of the $n$-th resonance with respect to the two-body threshold $2m$.

It is instructive to consider the single-resonance case ($N=1$) first and concentrate on the near-threshold region only, so that a nonrelativistic form of the loop operator can be employed, 
\be
G_{\rm nr}(k)=\frac{1}{16\pi m}(\varkappa+ik),
\label{Pi}
\ee
where $\varkappa$ is the properly regularised and renormalised real part of the loop treated as a free (input) parameter. Its sign is not fixed and it is only assumed that $|\varkappa|\ll m$ --- for the setting
studied here, where doubly heavy $\bar QQ$ states couple to a pair of heavy light states,
this relation emerges naturally since the size of the real part of the loop is predominantly 
set by the light quark physics. Then, after straightforward algebraic transformations, one finds that the scattering amplitude (\ref{eq1}) takes the form of the Flatt{\'e} distribution, introduced in Eq.~(\ref{Flatte}), with $E=k^2/m$ and the parameters
\be
E_f=\frac{\Delta_0^2}{m}-\frac12g_f\varkappa,\quad g_f=\frac{g^2}{8\pi},\quad\Delta_0^2=\frac14(M_0^2-4m^2),
\ee
where $M_0$ and $g$ are the bare mass of the resonance and its coupling to the field $\varphi$, respectively.

In the weak coupling regime ($g\to 0$) the amplitude possesses two symmetric poles in the complex momentum plane, $k_{1,2}=\pm\Delta_0$, both located either on the real or imaginary axis, depending on whether the bare resonance appears above ($M_0>2m$) or below ($M_0<2m$) the threshold. These symmetric poles describe a compact quark state, in agreement with the original setup. On the contrary, in the strong coupling limit ($g\to\infty$), 
the amplitude possesses only one pole located at $k_0=i\varkappa$ which, according to the pole counting rules of Ref.~\cite{Morgan:1992ge} and in line with the Weinberg picture introduced
above, corresponds to a molecule. Depending on the sign of $\varkappa$, this is either a bound or virtual state. Its formerly present counterpart pole has left the near-threshold region and moved to $-i\infty$ along the imaginary axis in this limit.

Let us now turn to the multi-resonance case and investigate the properties of the solutions of Eq.~(\ref{eqgen}). For vanishing couplings, $g_n=0$, we start from a set of $2N$ symmetric poles at $k_n^{(\pm)}=\pm\Delta_n$
which represent $N$ compact resonances. Depending on the values taken by the $\Delta$'s (real or imaginary), the corresponding poles reside either on the real or imaginary axis in the complex momentum plane.

Consider now the strong coupling limit. We will demonstrate now that there is one pole diving to $-i\infty$
in the $k$ plane in this regime, like in the single-resonance case considered above. This is a general feature of the system at hand, irrespective of the number of resonances and their original positions $\{\Delta_n\}$. Since, for the considered pole, $s\to-\infty$, then to study this case we obviously need to employ relativistic kinematics.
A convenient representation valid for real $s$ on the physical (thence superscript $^{(I)}$) sheet reads
\be
G^{(I)}_{\rm rel}(s)=\ds\frac{1}{16\pi^2 }
\left\{
\begin{array}{l}
\ds\frac{\pi\varkappa}{m}-r\ln\left|\frac{1+r}{1-r}\right|,\quad s<0\\[3mm]
\ds\frac{\pi\varkappa}{m} - 2r\ \mbox{arctan}\left(\frac{1}{r}\right),\quad 0<s<4m^2\\[3mm]
\ds\frac{\pi\varkappa}{m}-r\left(\ln\left|\frac{1+r}{1-r}\right|-i\pi\right),\quad 4m^2<s,
\end{array}
\right.
\ee
where $r=2|k|/\sqrt{\left|s\right|}$. In the nonrelativistic limit of $|k|\ll m$ the expression for $G_{\rm nr}$ from Eq.~(\ref{Pi}) is readily reproduced. 

In the opposite limit of $s\to -\infty$ we get 
\be
\lim_{s\to -\infty}G_{\rm rel}^{(I,II)}(s) = \mp\frac{1}{16\pi^2}\ln\left(-\frac{s}{m^2}\right),
\ee
where the upper and lower signs correspond to the loop operator evaluated on the first and second Riemann sheets indicated by the superscripts $(I)$ and $(II)$, respectively. 

To proceed with the argument we turn to the potential and make a natural assumption that the number of the below-threshold bare resonances is finite.
Moreover we take a value of the momentum sufficiently large to ensure that $|s|\gg |\Delta_1^2|$, where $\Delta_1$ corresponds to the lowest resonance in the spectrum. We do not need to assume a finite number of the bare above-threshold resonances since, in the considered limit, a remote above-threshold resonance $n$ contributes to the potential only a small (and gradually decreasing with $n$) amount $\propto 1/(-s+\Delta_n^2)$ with both $-s$ and $\Delta_n^2$ positive for the given kinematics. Moreover, it is natural to assume that the couplings $g_n$ decrease with $n$, so that more remote resonances are weaker coupled to the two-hadron channel. These conditions together allow one to truncate the sum in the potential at some sufficiently large number $n_{\rm max}$, neglect the contribution of all resonances with $n>n_{\rm max}$, and stick to the values of $s$ such that $|s|\gg \Delta_{n_{\rm max}}^2$. Then, in the considered limit, the potential reduces to
\be
V(s)\approx\left(-\frac{m^2}{s}\right)\sum_{n=1}^{n_{\rm max}}g_n^2>0,
\ee
where all $\Delta$'s were neglected compared with $s$. Then, for $-s\gg m^2$, the equation $V^{-1}(s)-G(s)=0$ takes the form
\be
\left(-\frac{s}{m^2}\right)=\mp\frac{1}{16\pi^2}\left(\sum_{n=1}^{n_{\rm max}}g_n^2\right)\ln\left(-\frac{s}{m^2}\right).
\label{eqs}
\ee

Equation (\ref{eqs}) possesses a sought solution with $-s\gg m^2$ only for the lower sign on the right-hand side, that is, for a virtual state. This solution cannot be presented in quadrature; however, its leading behaviour in the strong coupling regime is provided by the expression
\be
s\approx -\frac{m^2}{16\pi^2}\left(\sum_n g_n^2\right)\ln\left(\sum_n g_n^2\right),
\label{qlower}
\ee
which tends to infinity as the couplings increase. Importantly, such a solution exists for any $N$ (more precisely, for any $n_{\rm max}$), however, it is always unique. In particular, as argued above,
the pole which disappears to its infinitely far location is a virtual and not bound state pole.

Thus we have demonstrated on very general grounds that the number of poles of the $T$-matrix
gets reduced by 
one, when going from the weak coupling limit to the infinite coupling
limit. If we combine this nearly model-independent finding with the pole counting approach
we must conclude that the appearance of hadronic molecules in the spectrum is natural, however,
not necessary, only
in the strong coupling limit. We come back to this statement in Sec.~\ref{disclaimer}.

Let us now investigate the ultimate location of the other poles in the strong coupling limit, switching back to working in terms of the 3-momentum $k$. In this regime, the unity on the left-hand side (l.h.s.) of Eq.~(\ref{eqgen}) can be neglected, so that the pole positions come as solutions of the equation
\be
G(k)\sum_{n=1}^N \frac{g_n^2}{k^2-\Delta_n^2}=0
\ee
or, equivalently,
\be
G(k)\sum_{n=1}^Ng_n^2\prod_{
\begin{tabular}{c}
\\[-6mm]
\scriptsize $l=1$\\[-2mm]
\scriptsize $l\neq n$ 
\end{tabular}
}^N(k^2-\Delta_l^2)=G(k)P_{N-1}(k^2)=0,
\label{pol}
\ee
where $P_{N-1}$ is a polynomial of the order $N-1$ with real coefficient. 
Solutions of equation $P_{N-1}(k^2)=0$ form a set of $N-1$ pairs of symmetric poles
in the complex plane,
which represent dressed resonances, or bound states. Their positions appear somewhat shifted with respect to the original ones given by $\{\pm\Delta_n\}$. However, at least as long as we can assume
that the strong coupling regime is reached by 
\be
g_n\to g f(n), 
\label{gscaling}
\ee
where $g$ is
some universal factor and $f(n)$ accounts for the influence of the resonance label $n$
on the coupling strength, those pole locations will converge to certain locations with
a stable $g\to \infty$ limit. Note that a relation of the kind of Eq.~(\ref{gscaling})
emerges naturally when following the $N_c$ scaling of QCD, since then $g_n\propto 1/\sqrt{N_c}$.

In the meantime, the loop operator $G(k)$ provides an additional zero $k_0$ such that
\be
G(k_0)=0.
\ee
 As was argued
above, for heavy quark systems it appears to be natural that 
 $\varkappa$ is small compared to $m$. Then the nonrelativistic form of the loop operator from Eq.~(\ref{Pi}) can be employed, which gives
\be
k_0=i\varkappa.
\label{k0gamma}
\ee
Depending on the sign of $\varkappa$ the corresponding pole of the amplitude represents either a bound or virtual molecular state. 

The consideration presented above allows one to deduce a general pattern for the motion of the poles in the studied system, which holds irrespective of particular details of the model. Namely, starting from $N$ stable
states in the limit $g_n\to 0$ and coupling them to each other through the propagation of the $\varphi\bar{\varphi}$ pairs with an increasing coupling strength, we finally arrive at $N-1$ compact dressed states plus a molecular near-threshold state. It is instructive to estimate the values of the couplings necessary to approach the strong coupling regime when there appears the molecular pole. To this end we notice that,
away from the poles, the strong coupling regime implies that the product $V(k)G(k)$ is large compared with unity (see Eq.~(\ref{eqgen})). Then, for simplicity, considering all coupling of the same order $g$ and taking into account that (i) the denominator of the potential is of the order of the spacing between the resonances $\Delta^2_{n+1}-\Delta^2_n\simeq \Lambda_{\rm QCD}^2$, such that
$V\simeq m^2 g^2/\Lambda_{\rm QCD}^2$ and (ii) the loop operator takes values of the order $\varkappa/m$ (see Eq.~(\ref{Pi})), we arrive at the estimate
\be
g\gg\frac{\Lambda_{\rm QCD}}{\sqrt{m\varkappa}}\propto \frac{1}{\sqrt{M_Q}},
\label{est}
\ee
where the mass $m$ of the field $\varphi$ was substituted by the mass of the heavy quark $M_Q$. 

We conclude, therefore, that the appearance of near-threshold molecular hadronic states is a natural (imminent) consequence of the strong interactions between hadrons and, given the estimate (\ref{est}), it is easier to reach the limit of a sufficiently strong coupling in the spectrum of heavy quarks. In other words, we conclude that
\begin{itemize}
\item the spectrum of bottomonium may be rather rich in near-threshold exotic states; 
\item the considered mechanism may not apply to light quarks, where the strong coupling regime is much more difficult to reach.
\end{itemize}
 
\section{A concrete model}
 
We now turn to a more quantitative analysis. To avoid unnecessary technical complications, we start from 
coinciding coupling constants, $g_n=g$ for all $n$'s. For the masses of the resonances $R_n$ we stick to a linear dependence on the radial quantum number $n$ compatible with the linear confinement between quarks,
\be
M_n^2=(2M_Q)^2+\sigma n,\quad n=1,2,\ldots,N,
\label{Mn2}
\ee
where $\sigma$ is a parameter of the model of the dimension of mass squared. At the first step, we assume that all bare resonances reside above the threshold, $M_n>2m$. 

Then the evolution of the poles is described by the equation
\be
4m^2 G(s)\left[\sum_{i=1}^N \frac{1}{s-M_i^2}\right]=-\frac{1}{g^2},
\label{eq12}
\ee
and, for the spectrum (\ref{Mn2}), the sum over the resonances can be evaluated explicitly,
\bea
\sum_{i=1}^N \frac{1}{s-M_i^2}&=&\frac{1}{\sigma}\left[\psi\left(1-\frac{s-4M_Q^2}{\sigma}\right)\right.\nonumber\\[-2mm]
\label{pot2}\\[-2mm]
&-&\left.\psi\left(N+1-\frac{s-4M_Q^2}{\sigma}\right)\right],\nonumber
\eea
where $\psi(z)$ is the polygamma function, 
$$
\psi(z)\equiv\psi^{(0)}(z)=-\gamma+\sum_{i=0}^\infty\left(\frac{1}{i+1}-\frac{1}{i+z}\right),~\gamma\approx 0.57.
$$

In the weak coupling regime ($g\to 0$) Eq.~(\ref{eq12}) describes $N$ real poles $s_n^{(0)}=M_n^2$.
As the couplings $g$ deviates slightly from zero, the poles start to move into the complex $s$-plane, 
\be
s_n=M_n^2+\delta s_n,\quad |\delta s_n|\ll M_n^2,
\ee
and the equation for $\delta s_n$, as follows from Eq.~(\ref{eq12}), takes the form (for simplicity, only the leading, singular in $\delta s_n$ term is retained, which is sufficient for arguing)
\be
4m^2G(M_n^2)\left(\frac{1}{\delta s_n}-\frac{1}{\sigma}\bigl[\psi(N+1-n)-\psi(n)\bigr]\right)=-\frac{1}{g^2},
\label{dsnu}
\ee
where it was used that
\bea
\sum_{i=1}^{n-1}\frac{1}{s-M_i^2}&=&\frac{1}{\sigma}\bigl[\psi(n)+\gamma\bigr],\nonumber\\
\sum_{i=n+1}^{N}\frac{1}{s-M_i^2}&=&-\frac{1}{\sigma}\bigl[\psi(N-n+1)+\gamma\bigr],\nonumber
\eea
and the term with $i=n$ was considered separately.

The quantity $\psi(N+1-n)-\psi(n)$ has different signs for $n<(N+1)/2$ and $n>(N+1)/2$ and thus the expression in the square brackets in Eq.~(\ref{dsnu}) may change the sign, too. Then, since the sign of $\delta s_n$ crucially depends on the latter, there typically exists a boundary value $n_0$ such that 
all poles with $n<n_0$ are shifted to the right with respect to their original location for $g=0$ ($\mbox{Re}(\delta s_n)>0$) while all poles with $n>n_0$ are shifted to the left ($\mbox{Re}(\delta s_n)<0$). At the same time, the pole with the serial number $n_0$ is pushed away deep to the complex plane to travel a long path and thus to become a ``collective'' state --- this behaviour of the poles is discussed in Ref.~\cite{Hammer:2016prh}. 

Generalisation of the results obtained to more realistic versions of the model is straightforward. Namely, it is natural to assume that the coupling constants $g_n$ decrease with the serial number of the pole $n$, that is, the further the resonance resides from the threshold the weaker it couples to the hadronic channel. Meanwhile, from the consideration above it is easy to conclude that the appearance of the collective state depends on the combination $S_-(n_0)-S_+(n_0)$, with
\bea
S_-(n_0)=\sum_{n<n_0}\frac{g_n^2}{M_{n_0}^2-M_n^2},\nonumber\\[-2mm]
\label{Spm}\\[-2mm]
S_+(n_0)=\sum_{n>n_0}\frac{g_n^2}{M_n^2-M_{n_0}^2}.\nonumber
\eea

Therefore, changing the dependence of the couplings $g_n$ and masses $M_n$ on $n$, one can vary the serial number of the collective state $n_0$. However, from the discussion above it is clear that even for a scenario with the couplings decreasing extremely fast with $n$, still everything said above applies:
indeed, as demonstrated in Sec.~\ref{Weinberg}, even a single resonance turns into a molecular structure
in the coupling to infinity limit.

\begin{figure*}[t]
\centering
\raisebox{4mm}{\includegraphics[width=0.45\textwidth]{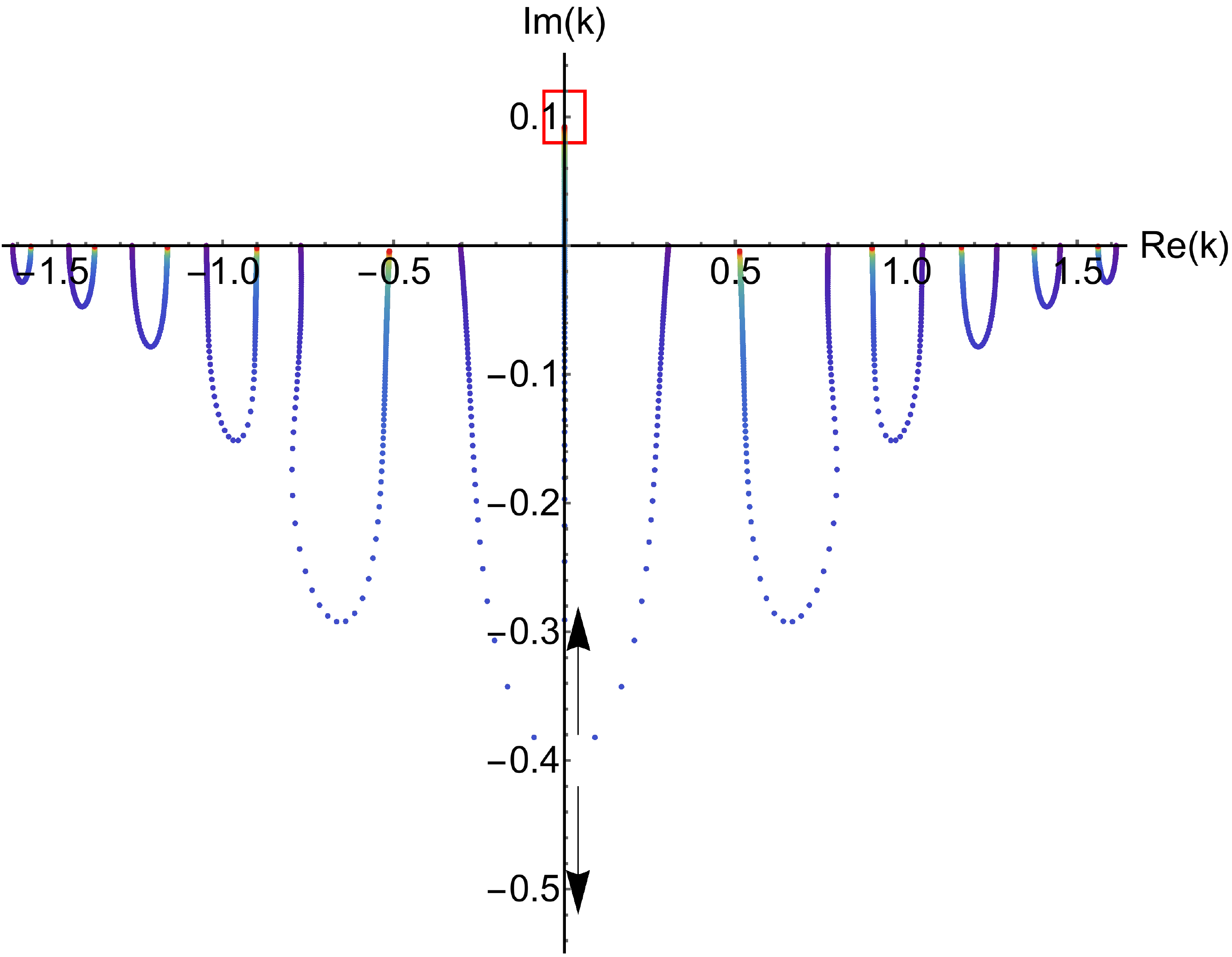}}\hspace*{0.05\textwidth}
\includegraphics[width=0.45\textwidth]{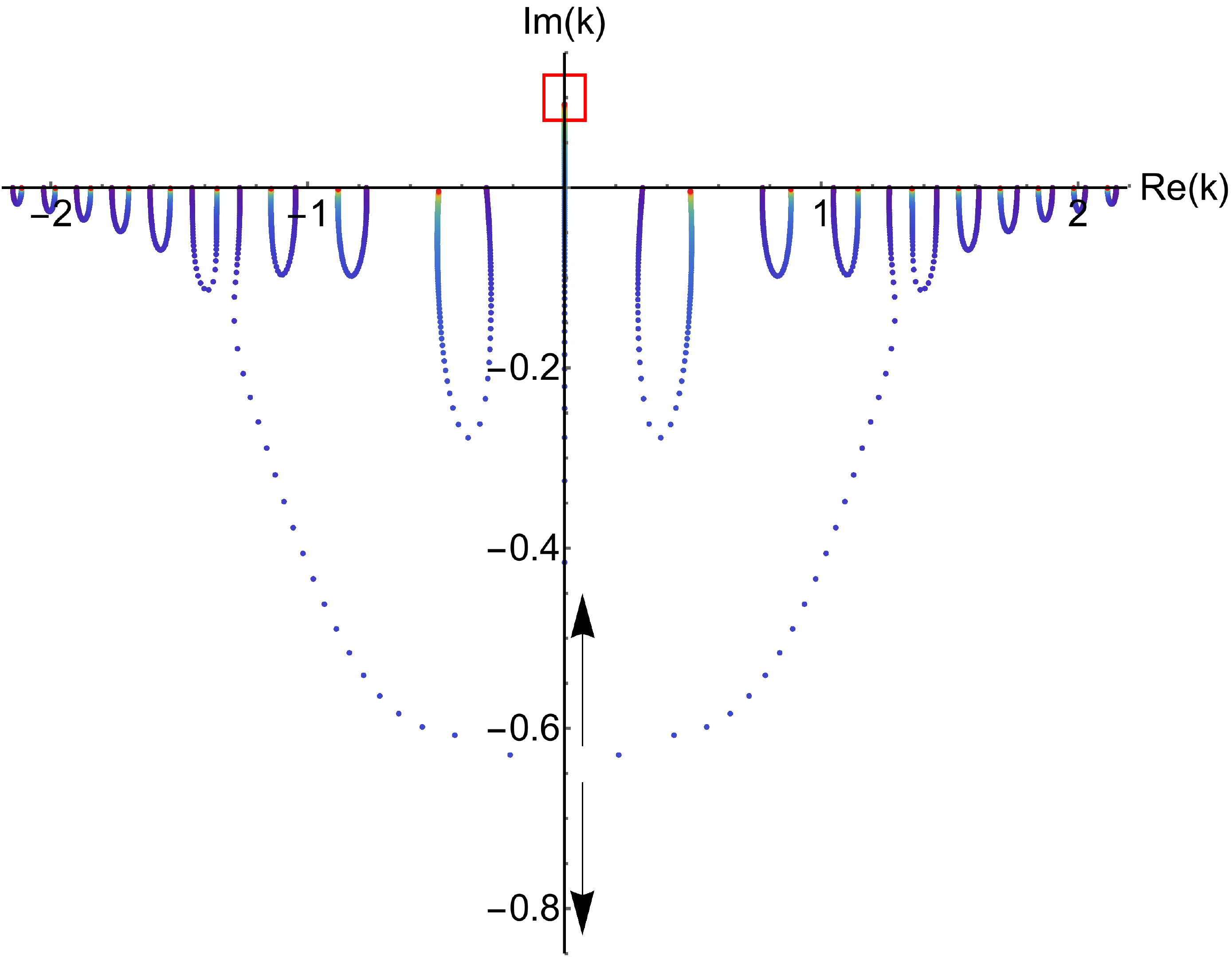}
\caption{The trajectories of the poles for $N=6$ (left plot) and $N=10$ (right plot) for model~A. The black arrows show the directions of the motion for the poles after they collide on the imaginary axis. The pole moving downwards dives to $-i\infty$ fast. The red square is centred at the point $k=i\varkappa=0.1$~GeV which is the final destination of the ``molecular'' pole in the limit $g\to\infty$.}
\label{fig:modelA}
\end{figure*}

\section{Numerical analysis}

After the qualitative analysis of the model (\ref{lag}) presented in the previous sections we now turn to a numerical investigation of the pole trajectories. In particular, for definiteness we set the parameters of the model as
\be
M_Q=2~\mbox{GeV},\quad \sigma=2~\mbox{GeV}^2.
\label{params}
\ee
Then the masses of the lowest resonances evaluated according to Eq.~(\ref{Mn2}) take the values 
$$
M_1=4.24~\mbox{GeV},~ M_2=4.47~\mbox{GeV},~ M_3=4.69~\mbox{GeV},~\ldots
$$

In the heavy-quark limit, the mass of the heavy-light meson $\varphi$ can be presented in the form 
\be
m=M_Q+\bar{\Lambda}+O(1/M_Q),
\ee
where $\bar{\Lambda}$ is a $M_Q$-independent constant related to the dynamics of the light quark, that is, $\bar{\Lambda}\simeq \Lambda_{\rm QCD}$. We consider two cases:\\[1mm]
\hspace*{1mm}$\bullet$ Model A:
\be
\bar{\Lambda}=0.1~\mbox{GeV}~\Longrightarrow~ m=2.1~\mbox{GeV}~(M_{\rm th}=4.2~\mbox{GeV})
\label{modelA}
\ee
\hspace*{1mm}$\bullet$ Model B:
\be
\bar{\Lambda}=0.2~\mbox{GeV}~\Longrightarrow~ m=2.2~\mbox{GeV}~(M_{\rm th}=4.4~\mbox{GeV})
\label{modelB}
\ee

It is easy to verify that in model~A $M_n>2m$ for all $n$'s while in model~B $M_1<2m$ and $M_n>2m$ for $n>1$. In other words, in the decoupling regime of $g=0$, all resonances reside above the threshold in model~A while, in model B, the lowest state appears below the threshold. In both models the loop operator is taken in the nonrelativistic form of Eq.~(\ref{Pi}). 

\begin{figure*}[t]
\centering
\raisebox{1.8mm}{\includegraphics[width=0.47\textwidth]{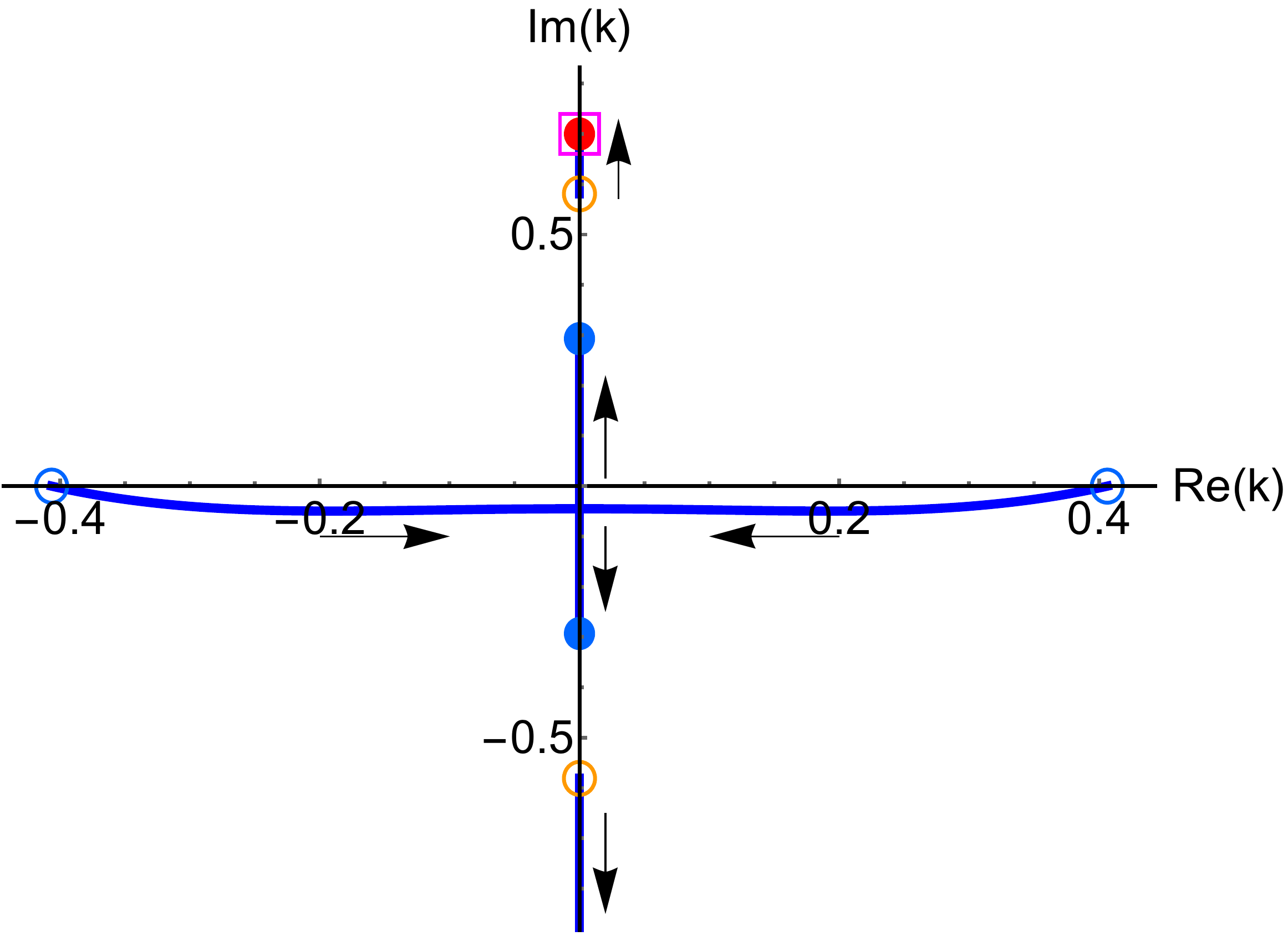}}
\includegraphics[width=0.47\textwidth]{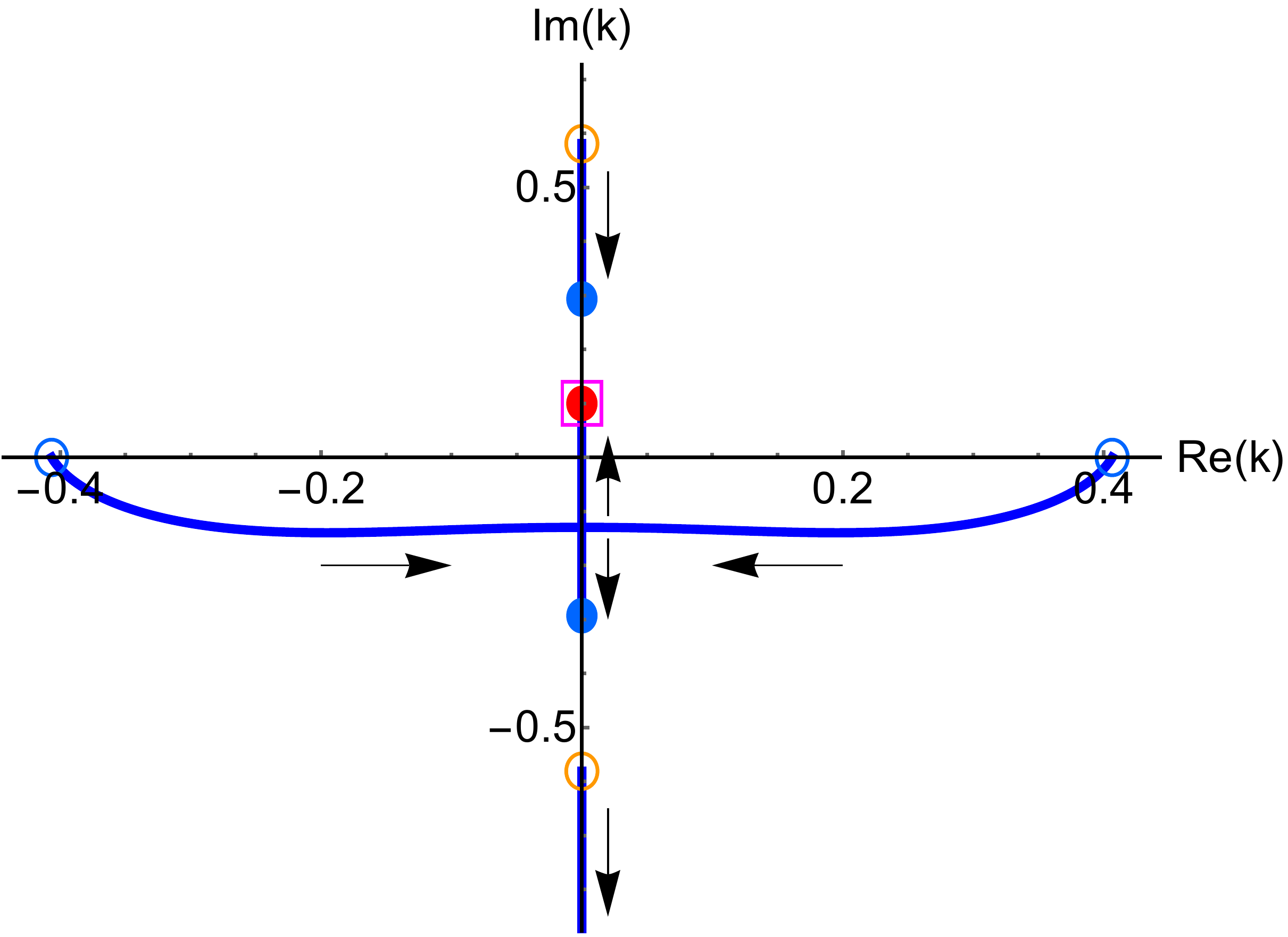}
\caption{The trajectories of the poles for $N=2$ for model~B. Left plot: $\varkappa>\sqrt{m^2-M_1^2/4}$ ($\varkappa=0.7$~GeV); right plot: $\varkappa<\sqrt{m^2-M_1^2/4}$ ($\varkappa=0.1$~GeV). The open circles of the same colour (orange and blue) show the pairs of symmetric poles which correspond to the bare resonances at $g=0$. The blue filled circles show the position of the poles which correspond to the dressed resonance in the strong coupling regime of $g\to\infty$. The red filled circle shows the position of the molecular pole in the strong coupling regime. The magenta square is centred at $k_0=i\varkappa$ to pinpoint the asymptotic position of the molecular bound state pole. The black arrows show the directions of the motion of the poles as the coupling $g$ grows.}
\label{fig:N2modelB}
\end{figure*}

\subsection{Model A}

The pole trajectories in the complex momentum plane for model A are visualised in Fig.~\ref{fig:modelA} --- for an illustration we choose $N=6$ and $N=10$. In the weak coupling regime, we start from $2N$ poles located symmetrically on the real axis. As the coupling grows, the poles get shifted to the complex plane and their trajectories start to bend. For $N=6$ (left plot in Fig.~\ref{fig:modelA}) all poles trajectories bend in the same direction while for $N=10$ (right plot in Fig.~\ref{fig:modelA}) the trajectories of the poles with $n<4$ and $n>4$ behave differently ($n_0=4$ corresponds to the collective state). Then, in the strong coupling regime (the red point at the end of each trajectory), $2(N-1)$ ``ordinary'' symmetric poles return back very close to the real axis and take positions between the original bare poles. On the contrary, the two poles for one selected resonance (collective state) travel a longer distance to hit each other on the imaginary axis in the lower half-plane. Then one pole from the pair dives fast towards the complex minus infinity (see Eq.~(\ref{qlower})) while the remaining one approaches its fixed final destination at $k_0=i\varkappa$ (see the red square centred at this point at each plot) which is the solution of the equation
\be
G(k_0)=\frac{1}{16\pi m}(\varkappa+ik_0)=0,
\ee
that is, it turns to a bound state with the binding momentum $\varkappa$ (the case of $\varkappa<0$ which corresponds to a virtual state looks similar and does not bring new understanding, so we do not discuss it here). According to the pole counting procedure discussed above, this single pole represents a molecular state. In other words, the behaviour of the poles for the collective state is identical to the behaviour of the poles in the single-resonance problem discussed above --- see Sec.~\ref{Weinberg}. 
In particular, in the strong coupling limit all other (compact) states have decoupled from the collective one, which naturally appears as a near-threshold hadronic molecule.
If this scenario is realised, it provides a justification for the studies
of isolated hadronic molecules, neglecting the influence from neighboring quark model states.

As one can see from Fig.~\ref{fig:modelA}, depending on particular model settings, the exceptional (collective) state is not necessarily the closest one to the threshold, like for the case of $N=6$ depicted in the left plot in Fig.~\ref{fig:modelA}. Indeed, changing the parameters of the model by, for example, increasing the number of the resonances $N$ (clearly, what matters is the balance between the partial sums $S_-$ and $S_+$ defined in Eq.~(\ref{Spm})), one can change the serial number of the state which turns to the molecule, as shown in the right plot in Fig.~\ref{fig:modelA}: The poles forming the resonance with $n=4$ travel long symmetric paths to bypass the trajectories of the poles with $n=1,2,3$. More detailed discussions on the emergence of the collective state can be found in Ref.~\cite{Hammer:2016prh}.
In the cited paper, also the dependence of the trajectory of the collective state on the 
parameters of $G(s)$ is discussed --- for the examples here we use the parameters quoted in Eqs.~(\ref{params}) and (\ref{modelA}), (\ref{modelB}).

Reverting the sign of $\varkappa$ results in a similar picture, however with a virtual state pole at $k=-i|\varkappa|$, so we do not discuss it in detail here. 

\subsection{Model B}

The motion of the poles in model~B demonstrates a certain similarity to that in model~A, however, with some subtleties. 
Thus, to avoid unnecessary complications, we consider the case of $N=2$ and the bare resonances with the serial numbers $n=1$ and $n=2$ located below and above the threshold, respectively. Then the following two cases need to be considered separately: $\varkappa>\sqrt{m^2-M_1^2/4}$ and $\varkappa<\sqrt{m^2-M_1^2/4}$ --- see the two panels of Fig.~\ref{fig:N2modelB}. In both panels the pole positions in the weak coupling regime are given by the open circles: orange and blue for the below- ($n=1$) and above-threshold ($n=2$) bare resonances, respectively, and the pole positions in the strong coupling regime are given by the three (two blue and one red) filled circles. 

Thus, for $\varkappa>\sqrt{m^2-M_1^2/4}$ (left panel of Fig.~\ref{fig:N2modelB}) in the limit of $g\to\infty$, the $n=1$ resonance turns to the molecule (the red filled circle) while the $n=2$ resonance gets dressed and moves below the threshold (the two symmetric blue filled circles), even closer to threshold than the
molecular state. On the other hand,
for $\varkappa<\sqrt{m^2-M_1^2/4}$ (right panel of Fig.~\ref{fig:N2modelB}) a kind of reordering of the poles takes place, for now the molecular pole (the red filled circle) originates from the $n=2$ resonance, while the two symmetric poles (the blue filled circles) for the dressed compact resonance come from different bare resonances. While in this case the pole trajectories are nontrivial, the outcome in the large coupling regime is as 
discussed in Sec.~\ref{sec:general}: In this limit we again arrive at one molecule and one compact state.

In case of a strongly fine-tuned system with $\varkappa=\sqrt{m^2-M_1^2/4}$ it is easy to verify that the bound state pole gets independent of the coupling constant $g$ and keeps its original position at $k_0=i\varkappa$ for all
values of $g$, while its counterpart starts at $k_0'=-i\varkappa$ and then, as $g$ grows, dives to $-i\infty$ fast, in agreement with the general considerations presented above.

Increasing the number of bare resonances and/or changing the number of below-threshold ones would not bring new insight since, in the strong coupling regime, the behaviour of the poles always follows the pattern outlined above, namely,
\begin{itemize}
\item if there are no bare below-threshold resonances, at least one of the above-threshold resonances (not necessarily the closest one to the threshold), after dressing, moves below the threshold;
\item the lowest pole lying on the imaginary axis in the lower half plane moves towards $-i\infty$;
\item one of the poles lying on the imaginary axis in the upper (lower) half plane approaches the zero of the loop operator at $i\varkappa$ ($-i|\varkappa|$) to represent a molecular bound (virtual) state with the binding momentum $\varkappa>0$ ($\varkappa<0$);
\item all other poles compose symmetric pairs to represent the dressed compact resonances --- whether the poles for a given dressed resonance stem from the same bare resonance or a rearrangement of poles takes place depends on the value of the parameter $\varkappa$ in relation with the bare pole positions.
\end{itemize}

\begin{figure*}[t]
\centering
\includegraphics[width=0.49\textwidth]{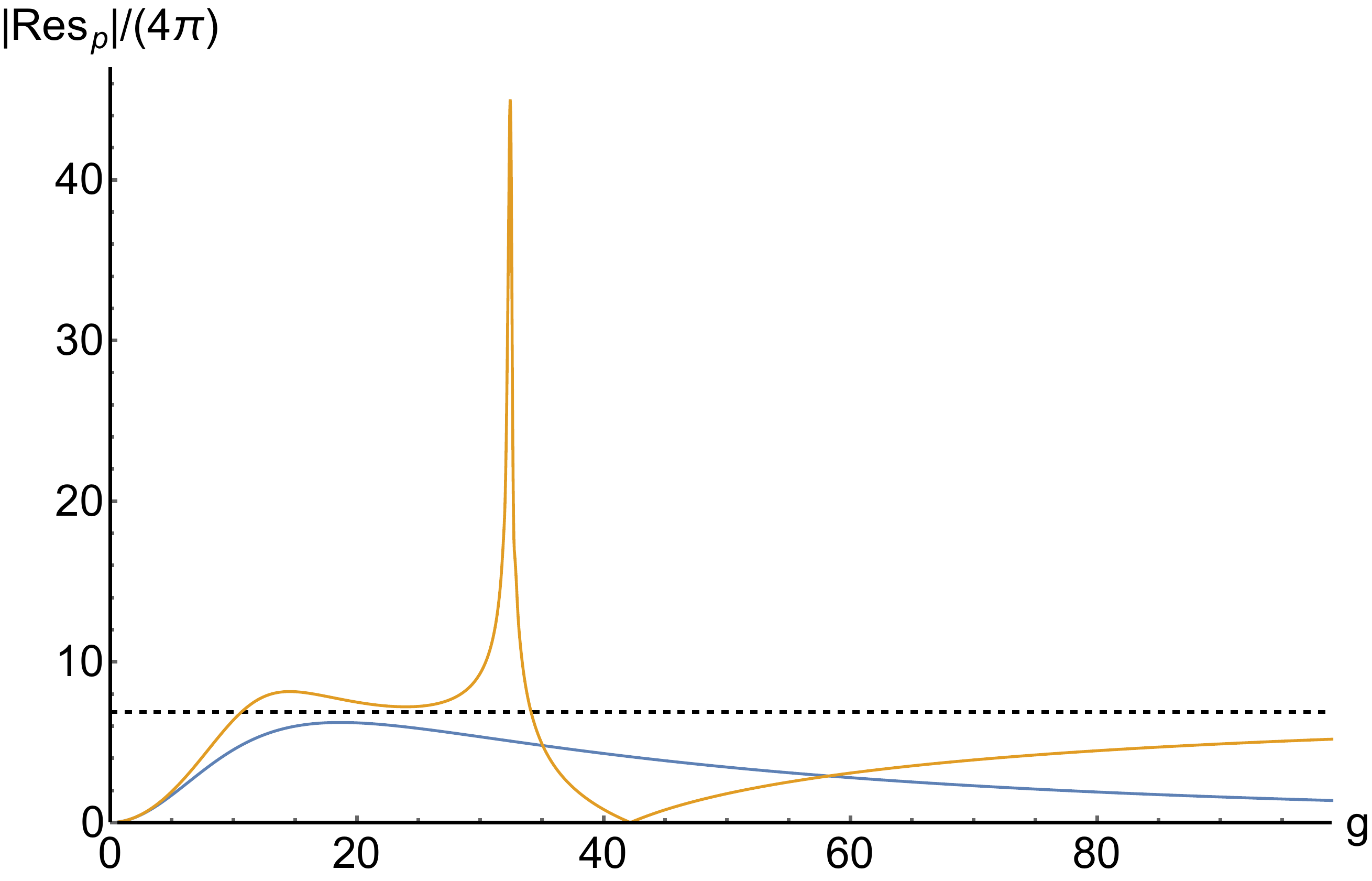}
\includegraphics[width=0.49\textwidth]{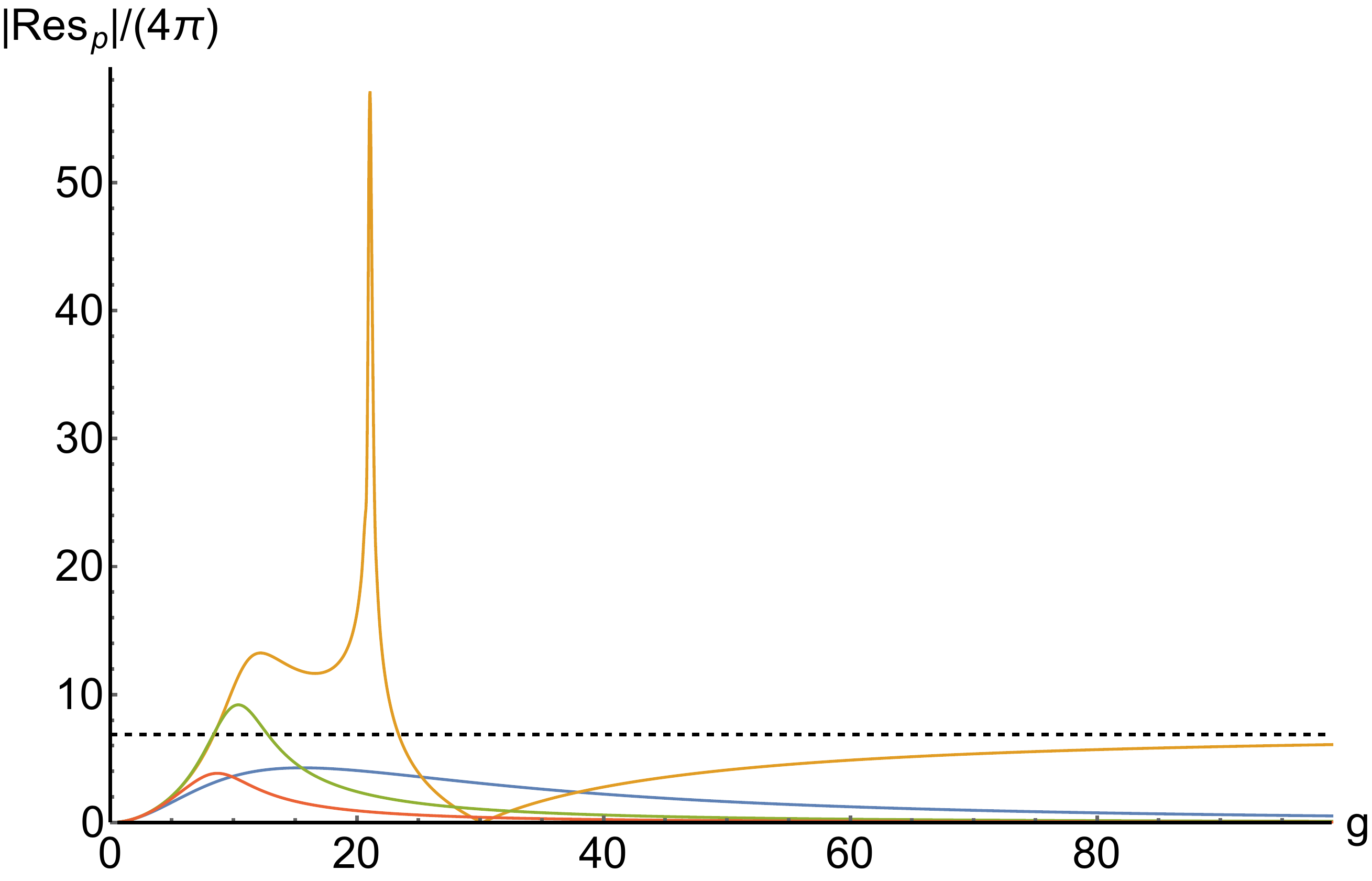}
\caption{The values of the $|\mbox{Res}_p|/(4\pi)$ at the poles (see Eq.~(\ref{resp})) as functions of the dimensionless coupling $g$ for model B with $\varkappa=0.1$~GeV and $N=2$ (left panel) and $N=4$ (right panel). The blue curve is for the below-threshold pole with $n=1$. The orange curve which represents the collective state with $n=2$ asymptotically approaches the limiting value quoted in Eq.~(\ref{Res0}) (shown with the horizontal black dashed line). The green and red curves in the right panel are for the poles with $n=3$ and $n=4$, respectively.}
\label{fig:res}
\end{figure*}

\section{Residues}
\label{residues}

In the vicinity of the pole $s_p$, such that
\be
V^{-1}(s_p)-G(s_p)=0,
\label{pole}
\ee
the amplitude takes the form
\bea
T(s)&=&\frac{1}{V^{-1}(s)-G(s)}\nonumber\\
&\ds\mathop{\approx}_{s\to s_p}&\frac{1}{\left.(d/ds)[V^{-1}(s)-G(s)]\right|_{s=s_p}(s-s_p)}\\
&=&\frac{\mbox{Res}(s_p)}{s-s_p}.\nonumber
\eea
We, therefore, have
\be
\mbox{Res}(s_p)=\frac{8k_p}{(d/dk)\left.[V^{-1}(k)-G(k)]\right|_{k=k_p}},
\ee
where we used $s=4(k^2+m^2)$ to switch from the derivative with respect to $s$
to that with respect to $k$. Using the explicit form of the propagator
provided in Eq.~(\ref{Pi}) as well as the relations (\ref{pole}) and (\ref{gscaling}), this can be written as
\bea
\mbox{Res}_p&\equiv&\mbox{Res}(s_p)=-\frac{8k_p}{G'(k_p)+V^{-2}(k_p)V'(k_p)}\nonumber\\[-2mm]
\label{resp}\\[-2mm]
&=&-\frac{8k_p}{i/(16\pi m)+g^2G^2(k_p)(V'(k_p)/g^2)}.\nonumber
\eea
As argued below Eq.~(\ref{gscaling}), it is natural to assume that $(V'(k_p)/g^2)$ is independent
of $g$. Thus, in Eq.~(\ref{resp}) the $g$-dependence of the residue is explicit and
we can straightforwardly discuss the
scaling of the residues in the large-coupling limit. 

The collective pole emerges from the zero of $G(k)$, $k_0\to i\varkappa$, in the strong coupling regime of $g\to \infty$\footnote{Clearly here one first has to approach the pole and then take
the $g\to \infty$ limit, following the pole.}. Therefore, in this limit, the amplitude takes the form (see the discussion in Sec.~\ref{Weinberg})
\be
T(s_0)=-\frac{g_{\rm eff}^2}{s-s_0},\quad s_0=4(m^2+k_0^2),
\ee
with
\be
\frac{g_{\rm eff}^2}{4\pi}=32\varkappa m,
\label{Res0}
\ee
which agrees to the Weinberg's universal coupling --- see Eq.~(\ref{geffdef}) with $\lambda=0$ (pure molecule) and $m_1=m_2=m$ which gives $\mu=m/2$.

For all other poles $G(k_p)\neq 0$, so that the leading dependence of the residue (\ref{resp}) on the coupling constant appears to be
\be
\mbox{Res}_p\propto \frac{1}{g^2}.
\ee

The pole positions depend on $g$ as well, however, all poles related to the nonmolecular
states approach a well defined location in the large coupling limit. Therefore,
this dependence does not change the general pattern that the residues of all 'ordinary' poles decrease
as $1/g^2$ in the strong coupling regime.

The behaviour of the residues for model B with $\varkappa=0.1$~GeV is exemplified in Fig.~\ref{fig:res}. For the parameters used the asymptotic value of the residue for the collective state calculated as given in Eq.~(\ref{Res0}) equals 6.88 GeV$^2$ (see the horizontal black dashed line in Fig.~\ref{fig:res}).

\section{Disclaimer and Discussion}
\label{disclaimer}

While the properties of the model outlined here appear to emerge very generally
when the coupling parameter is varied, we should stress that it is still a model. 
The only feature that is solidly nested within QCD is the weak coupling regime that
is reached in the large-$N_c$ limit, as already mentioned above.
The large-$N_c$ limit is known to provide an idealised but quite instructive limit for QCD which shares many important features of the theory realised in nature with $N_c=3$. 
In this limit, the coupling of a quarkonium to a pair of mesons scales as $1/\sqrt{N_c}$ and vanishes as $N_c$ grows. This provides a natural realisation of the weak coupling regime
where an infinite tower of stable states with the $N_c$-independent masses appears --- for
the reasoning here it does not matter if those are $\bar QQ$ mesons or more
complicated structures like tetraquarks which might also survive the large-$N_c$ limit~\cite{Weinberg:2013cfa,Maiani:2018tfe,Lucha:2021mwx}.
However, as one starts to reduce $N_c$ thus increasing the coupling, not
only start the poles to talk to each other in the way outlined here via simple
meson loops, but also $t$- and $u$-channel meson exchanges between hadrons involved become possible
introducing additional scales into the problem. 
Moreover, in the simple scheme outlined here the two-hadron loops $G(s)$ come with the same sign at all energies and thus the resonance potentials all add coherently,
which eventually drives the emergence of the collective state. In more
realistic settings, where also $t$- and $u$-channel exchanges
are present, this coherence can get spoiled --- had we formulated the
model in terms of meson exchanges, in this scenario we would have a repulsive
potential. Then clearly no collective state gets generated. On the other hand,
in all those cases where the emerging meson exchanges do not spoil the 
coherence, it follows from the consideration of this paper that the emergence
of the collective state or hadronic molecule is very natural. Still, the 
meson exchanges leave an imprint in the results.
The implications of this observation are most
easily explained by their impact on Eq.~(\ref{Res0}): In the scenario discussed in this paper
this equation is exact in the infinite coupling limit. In a more realistic model the same
relation emerges; however, it is subject to corrections that scale either as the
binding momentum times the range of forces or the biding momentum divided by the distance
to the closest threshold~\cite{Matuschek:2020gqe}.

Another comment is also in order: Quark-hadron duality tells us that an infinite sum of $s$-channel
poles can but does not have to map onto an infinite sum of $t$-channel poles. 
Therefore our study here does not allow
for any conclusions on the binding mechanism. It does not even imply that there needs
to be an infinite tower of $s$-channel poles present in the large-$N_c$ limit, for $t$-channel
exchanges can still be operative (and bind) even in the absence of those.

What this paper does provide, however, is a mechanism that connects the 
large-$N_c$ limit of QCD with a scenario in the real world where hadronic molecules
naturally emerge and decouple from the surrounding quark model states. If this scenario is
indeed realised in nature, if provides a justification to investigate hadronic molecules
independently from compact quark-model states that might or might not exist in their
neighborhood. 
 
\section{An option for the $D_{s1}(2460)$}

\begin{figure*}[t]
\centering
\includegraphics[width=0.47\textwidth]{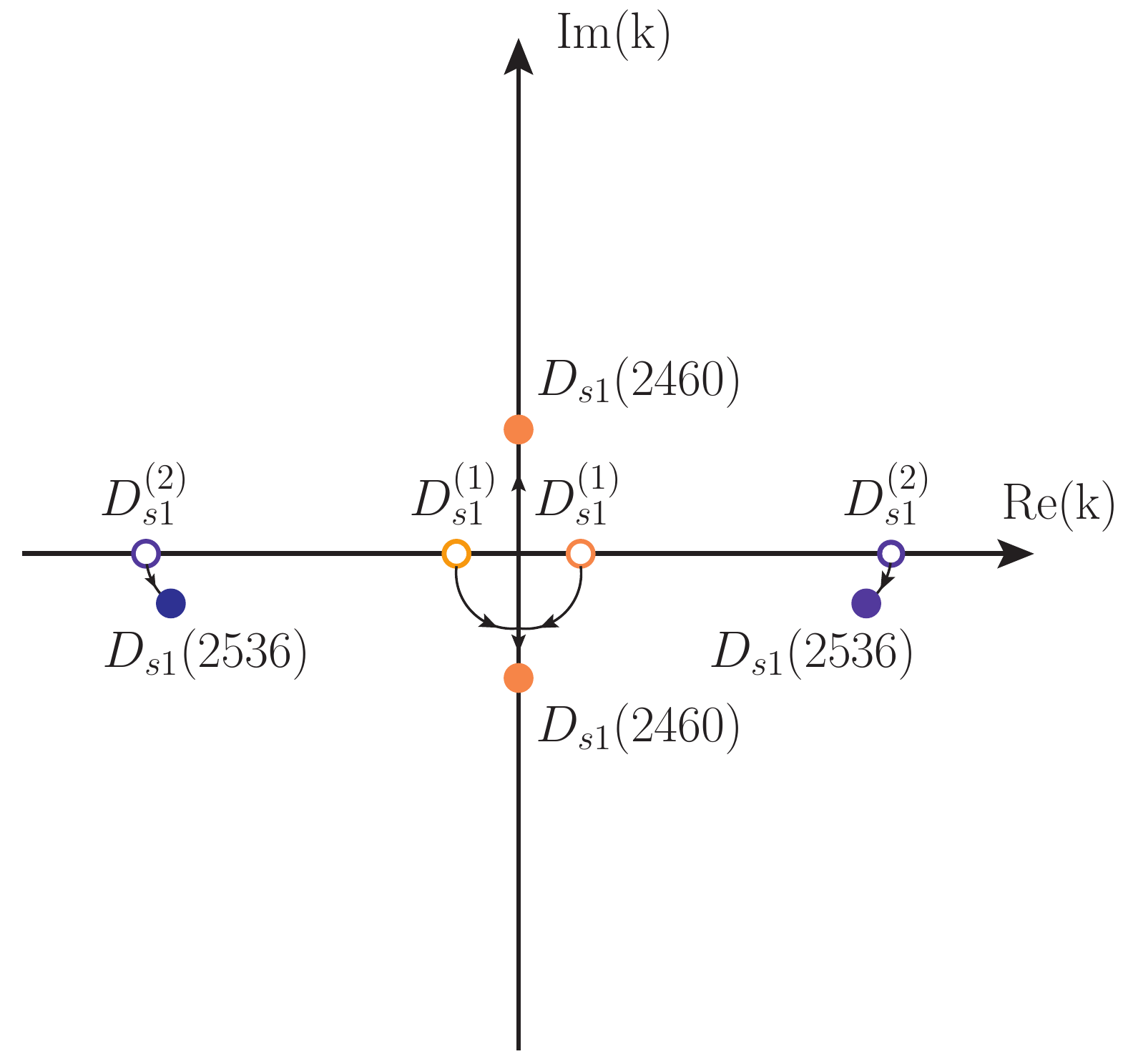}
\includegraphics[width=0.47\textwidth]{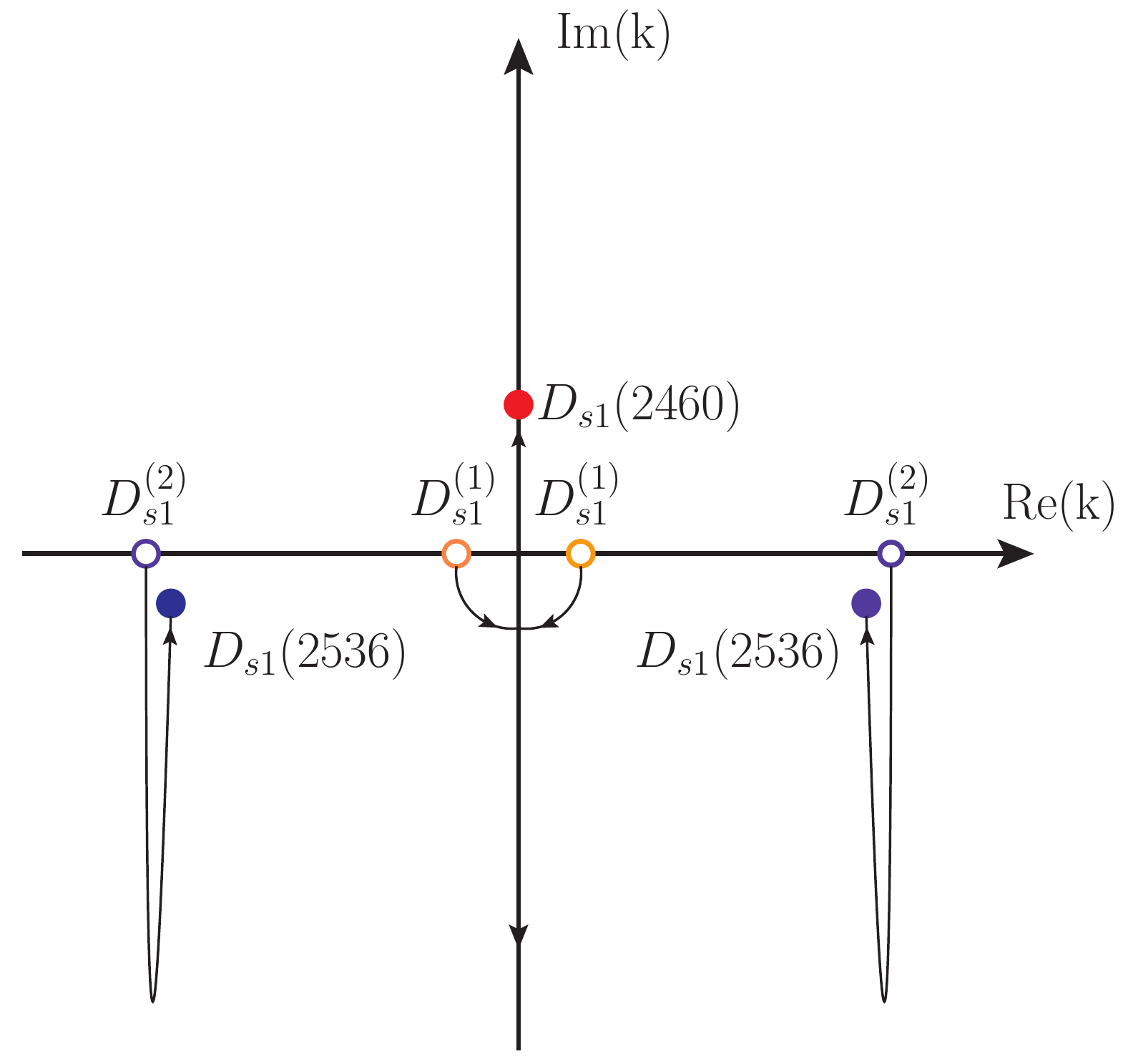}
\caption{Two scenarios for the $D_{s1}(2460)$. The weak coupling regime (left plot) and strong coupling regime (right plot). The open (filled) circles show the positions of the bare (dressed) resonances. The $D_{s1}(2460)$ is a compact quark state (two nearly symmetric orange circles in the left plot) in the weak coupling scenario and a molecule (single red filled circle in the right plot) in the strong coupling regime --- see the text for the details. The black arrows show the poles motion.}
\label{fig:ws}
\end{figure*} 

An obvious question to ask now is what would be a signature of the scenario 
discussed above in the hadron physics phenomenology. The prediction is that in a channel
where there is a hadronic molecule and at the same time compact quark states, 
the latter should (largely) decouple from the channel that forms the molecule. 
A promising system that one can confront with this prediction is provided by the strange
charm mesons with $J^P=1^+$. Indeed, in this channel not only a promising candidate for the $D^*K$ molecule --- the $D_{s1}(2460)$ --- exists which lies only 42 MeV below the corresponding threshold,
but also an additional state with the same quantum numbers lies somewhat higher up in the spectrum, namely the $D_{s1}(2536)$. 
As we discuss below in this section, the properties of these two states can be well described 
in both weak and strong coupling scenarios, although the former requires some fine tuning
while the latter emerges naturally.

To set the stage let us start from the quark model.
Then, for an arbitrary heavy-quark mass $M_Q$, the doublet of the physical observed states with the quantum numbers $J^P=1^+$, denoted below as $P_1^l$ and $P_1^h$
for the light and heavy members of the doublet, respectively, 
come as particular combinations of the $\{^{2S+1}L_J\}$ basis vectors $^1P_1$
and $^3P_1$,
\be
\left(P_1^l\atop P_1^h\right)=
\left(
\begin{array}{cc}
\cos\theta(M_Q)&-\sin\theta(M_Q)\\
\sin\theta(M_Q)&\cos\theta(M_Q)
\end{array}
\right)
\left({}^1P_1\atop {}^3P_1\right).
\label{thetaP}
\ee
The mixing originates from spin-dependent terms in the Hamiltonian of the $Q\bar{q}$ ($q\bar{Q}$) meson and, in the strict limit of $M_Q\to\infty$, is described by the ideal mixing angle $\cos(\theta(\infty))=1/\sqrt{3}$. In this limit, the physical states correspond to the total momentum of the light degree of freedom $\vec{j}=\vec{l}+\vec{s}$ equal to $j=1/2$ and $j=3/2$. For the physical $c$-quark mass, however, given that $\Lambda_{\rm QCD}/m_c\sim 0.3$, the mixing angle should deviate from its ideal value. As a result, the $D_{s1}(2460)$ and its counterpart $D_{s1}(2536)$
are expected to be particular combinations of the $1^1P_1$ and $1^3P_1$ quark-antiquark states. 
In this scenario the mixing angle may be fixed from the width of the $D_{s1}(2536)$ which
decays predominantly to the $D^*K$ pair. In Ref.~\cite{Song:2015nia} it is found that the width of the $D_{s1}(2536)$ is consistent with the data \cite{ParticleDataGroup:2022pth},
\be
\Gamma[D_{s1}(2536)]=0.92\pm 0.05~\mbox{MeV},
\label{Ds1tot}
\ee
only if the mixing angle is very close to the ideal one.

It is easy to see that the scenario described in Ref.~\cite{Song:2015nia} corresponds to the regime of a weak coupling with the continuum channel $D^*K$. In this regime the poles get only slightly shifted from their original positions to represent the physical states $D_{s1}(2460)$ and $D_{s1}(2536)$. 
Thus in this scenario the quark model, without couplings to the continuum, should already
produce states in a close vicinity of their physical pole positions.
However, as soon as the mixing angle deviates from the ideal one, which should be expected in 
the charm sector,
the partial decay width $D_{s1}(2536)\to D^*K$ grows fast to take values $\sim 10$~MeV~\cite{Song:2015nia}, an order of magnitude larger than the experimental $D_{s1}(2536)$ total width (\ref{Ds1tot}). 
Thus, the small $D_{s1}(2536)$ width comes in this scenario as a result of a certain fine tuning for the mixing angle. 
In this scenario, both $D_{s1}(2460)$ and $D_{s1}(2536)$ are compact quark states. This scenario is visualised in the left panel of Fig.~\ref{fig:ws}.

The strong coupling regime as proposed in this work can provide an alternative explanation
for the properties and structure of the two $D_{s1}$ states. As before, we start from the bare quark resonances $D_{s1}^{(1)}$ and $D_{s1}^{(2)}$. At least one of them should appear above the threshold while no constraint is imposed on the position of the other one. As the coupling with the continuum channel $D^*K$ grows, the width of the upper state increases first and then starts to decrease to approach some small value provided the coupling is sufficiently large. The poles in the complex momentum plane, which represent this state, always remain symmetric, so that the physical $D_{s1}(2536)$ meson survives as a compact quark state, however, with an effective coupling to $D^*K$ much smaller than expected
by the quark model. In the meantime, the poles representing the lower state behave differently: as explained above, one of them disappears from the near-threshold region while the other one approaches a certain position defined by the properties of the $D^*K$ system. Thus, in this strong coupling scenario, the physical $D_{s1}(2460)$ state appears to be a molecule --- see the right panel in Fig.~\ref{fig:ws}. 
 
A comment on how the strong-coupling regime can be reached in this system appears helpful here. Employing the estimate (\ref{est}) with the reduced mass of the $D^*K$ system $m\approx 400$~MeV and the binding momentum $\varkappa=\sqrt{2m E_B}\approx 180$~MeV, with $E_B=m_{D^*}+m_K-m_{D_{s1}(2460)}\approx 40$~MeV one finds the critical value of the coupling of the order unity. In other words, the dimensionless coupling constant of the natural size $g\simeq 1$ would already provide a dynamics of the system compatible with the strong coupling regime. As explained above, in this case the counterpart of the bound state pole appears far away from the near-threshold region, and the compact component of the resonance wave function is suppressed compared with its molecular component. Meanwhile, the simple model employed in this paper does not allow for more quantitative conclusions and is only aimed at providing a qualitative picture of the phenomenon.
 
But how can one distinguish between the two scenarios experimentally?
Here the most striking signature is predicted to come from the
hadronic decay width of the $D_{s1}(2460)$: since its mass is below
the $D^*K$ threshold, its only allowed strong decay is to $D_s^*\pi^0$.
This decay violates the conservation of isospin and is thus expected
to be rare. Indeed, in a scenario where the positive-parity
charm-strange states are treated to be $c\bar s$ mesons, the 
hadronic width that emerges is of the order of 10 keV~\cite{Colangelo:2003vg}\footnote{The calculation presented in that
paper is performed for the $J=0$ partner state of the $D_{s1}(2460)$,
namely the $D_{s0}(2317)$; however, heavy quark spin symmetry
makes one expect the width of the axial vector state to be
very similar.}
However, if the $D_{s1}(2460)$ is a hadronic molecule, its hadronic
width is significantly enhanced by $D^*K$ loops. Indeed, the coupling to this channel is large for the molecule and the isospin violation is enhanced since the mass splitting of the thresholds for the two channels contributing to the isoscalar $D_{s1}(2460)$ state, the
$D^{*+}K^0$ and $D^{*0}K^+$, are of the same order of magnitude as
the binding energy of the state~\cite{Lutz:2007sk,Guo:2008gp,Liu:2012zya,Cleven:2014oka,Guo:2018kno,Fu:2021wde} --- in
fact, this is the same kind of mechanism that enhances the mixing of the light scalars $a_0(980)$
and $f_0(980)$ if those states are treated as hadronic molecules~\cite{Achasov:1979xc,Hanhart:2007bd}.
For example, Ref.~\cite{Fu:2021wde} quotes the $D_{s1}(2460)$ width
as large as $(111\pm 15)$~keV --- admittedly a challenge for the experiment,
but an order of magnitude larger than the prediction for the quark-antiquark structure. 

\section{Summary}

In this work we presented a general description of the motion of the poles in a system of $N$ compact hadronic resonances interacting through their coupling to a two-meson state. 
In particular, we start from a set of $2N$ symmetric poles in the complex momentum plane which represent the bare resonances $R_n$ (heavy quarkonia) and then couple them to a scalar field $\varphi$ (a scalar meson containing a heavy quark and a light antiquark). Some bare poles appear below and some above the $\varphi\bar{\varphi}$ threshold. In the regime of small coupling, the poles lying above the threshold get shifted to the complex plane and then, as the coupling increases, their trajectories bend and reapproach the real axis. Such a behaviour of the poles was previously discussed in the literature --- see, for example, Refs.~\cite{vanBeveren:2006ua,Hammer:2016prh,Ortega:2021fem}. Although, naively, it may look unnatural, there are good physical reasons for it to be true. Indeed, a strong coupling to the continuum channels tends to increase the width of the resonances, however the strong unitarisation effects which play an important role in this regime do not allow the poles to move to the complex plane far away from the real axis. Eventually the unitarisation effects win and the poles move back towards the real axis. As a result, despite multiple open decay channels and a large phase space available for the decays of the resonances into these channels in the strong coupling regime, excited hadrons do not turn into extremely broad and strongly overlapping humps, but should survive as relatively narrow structures in the spectrum --- at least in the heavy quark sector. In some cases, above-threshold poles can move below the threshold as a result of the strong interaction with the field $\varphi$. The poles lying below the threshold always move along the imaginary axis. 

In the strong coupling regime, we arrive at $2(N-1)$ symmetric poles in the complex momentum plane representing $N-1$ compact dressed quarkonia which may lie both below or above the threshold. In the meantime, the remaining pair of poles behaves differently --- although, in the weak coupling regime, they also correspond to compact resonances, in the strong coupling regime, one of them leaves the near-threshold region and tends to $-i\infty$ in the momentum plane, while the other one approaches a fixed point $i\varkappa$ provided by the zero of the loop operator for the field $\varphi$. Depending on the sign of $\varkappa$, this is either a bound or virtual state; however in either case it qualifies as a molecule. Interestingly, the fate of the molecular pole in the strong coupling regime is defined by the properties of the loop operator evaluated for the free field $\varphi$, that is, a kind of duality between the strong and weak coupling regimes takes place. We exemplify our finding by simple model calculations.

In the picture drawn in this work near-threshold molecular states appear naturally and, furthermore, cannot be avoided provided the coupling of the quark resonances to the continuum channel is strong enough (the strong coupling limit is reachable) and the coherence of the effect of the different 
resonances does not get spoiled by other effects like $t$-channel
exchanges that also contribute as soon as the coupling gets large. We find that the critical value of the coupling needed to reach the strong coupling regime is inversely proportional to the mass of the quark, so that approaching the strong coupling regime, which is unlikely for light quarks, appears to be plausible in practice for heavy quarks. In particular, our picture favours a rather rich family of near-threshold exotic states in the spectrum of bottomonium.

We demonstrate explicitly how the molecular state, which appears as a result of the strong coupling of the compact quark resonances with a continuum channel, can coexist with compact (dressed) quark resonances located both below and above the threshold. Moreover, if the dressed above-threshold resonances exist, specific predictions for them can be made: since the trajectories of the poles for the above-threshold poles do not continue deep to the complex plane, when the coupling increases, but bend such that the poles return back to the real axis, the strong coupling regime entails ``unnaturally'' small widths of the dressed quarkonia, which may appear at odds with the predictions of the quark model. 
In other words, we find that the coexistence in the spectrum of heavy quarks of a near-threshold molecular state and an unnaturally narrow above-threshold quark state(s) with the same quantum numbers may signal our proposed mechanism at work.
We confronted the properties of the lowest charm-strange $J^{p}=1^+$ states
with this picture and found consistency, although also other explanations
are possible. A straightforward conformation of our scenario would be
if the hadronic width of the $D_{s1}(2460)$ were found to be about 100~keV or above.

\begin{acknowledgments}
This work is supported in part by the Deutsche Forschungsgemeinschaft (DFG) 
through the funds provided to the Sino-German Collaborative Research Center ``Symmetries and the Emergence of Structure in QCD'' (NSFC Grant No. 12070131001, DFG Project-ID 196253076 -- TRR110). A.N. is supported by the Slovenian Research Agency (research core funding No. P1-0035).
\end{acknowledgments}

\end{document}